\documentclass[twoside]{article}
\usepackage{fleqn,espcrc2,times}
\usepackage{graphicx}
\graphicspath{/home/ykao/PAPERS/M2S/Figures}%
\DeclareGraphicsExtensions{.eps,.ps}
\newcommand{\AmS}{{\protect\the\textfont2
  A\kern-.1667em\lower.5ex\hbox{M}\kern-.125emS}}

\title{Commensurate and
 Incommensurate Structure of the Neutron Cross Section
in \mbox{LaSrCu} and \mbox{YBaCuO}}
  \author{Ying-Jer Kao 
  \address{James Franck Institute, The University of Chicago, 5640
    South Ellis Avenue, Chicago, Illinois 60637}
  ,Qimiao Si
  \address{Department of Physics, Rice University, Houston, TX 77251}
  ,
and K. Levin$^{\rm\  a}$ 
  }

\begin{document}
%----------------------------------------------------------------

\begin{abstract}
  We study the evolution of the $d$-wave
neutron cross-section with variable
  frequency $\omega$ and fixed $T$ (below and above $T_c$)
 in two different cuprate
  families. The evolution from incommensurate to commensurate to
incommensurate peaks is rather generic within an RPA-like scheme.
This behavior seems to be in reasonable accord with experiments,
and may help distinguish between this and the ``stripe" scenario.
\end{abstract}

\maketitle
%------------------------------------------------------

The goal of the present paper is to address the frequency evolution
of the neutron cross section, over the entire range of $\omega, {\bf q}$, 
using
a scheme which we have previously applied to the normal\cite{Si} and
$d$-wave superconducting states\cite{Zha,Liu}. 
This work is viewed as significant because it
leads to a fairly generic frequency evolution, which seems to be
observed experimentally in two cuprate families\cite{Mook,newMook,Japanese,Keimer}.  
These calculations, which have no adjustable
parameters (besides those which were used to fit the normal state), 
can help establish
whether the details of the fermiology plus $d$-wave pairing can account
for the observed incommensurabilities and their evolution with
frequency, or whether (by default) some new phenomenon such as stripe
\cite{stripes} or other exotic\cite{SO5} phases may be required.

Our starting point is a three band, large $U$ calculation\cite{Si} which
yields a dynamical susceptibility
 $ \chi ( q , \omega ) = \chi ^o ( q , \omega ) / [ 1 + J ( q ) \chi ^o ( q ,
  \omega )]$
where the Lindhard function $\chi^o$ is appropriate to the ($d$-wave)
superconducting state and the underlying Fermi surface\cite{Zha,Littlewood,Tremblay,Bulut}.  
Here the residual exchange $J(q) = J_o [\cos q_x +\cos q_y] $ arises
from Cu-Cu interactions via the mediating oxygen band.
While the YBaCuO system is a two layer
material, our past experience has shown that most of the peak structures
associated with the neutron cross section are captured by an effective
one layer band calculation, which we will investigate here.
For definiteness, we fix the temperature at $4$ K and vary
frequency in increments of a few meV.  We take the electronic excitation
gap to be described by an ideal $d$-wave, $\Delta ( q ) = \Delta (\cos
q_x +\cos q_y)$, where at $ T = 4 $ K, $\Delta$ is taken to be $17 $ meV 
in YBaCuO$_{6.6}$ and $8 $ meV in optimally doped LaSrCuO.  

\begin{figure*}
\centerline{
\includegraphics[width=6.5in]{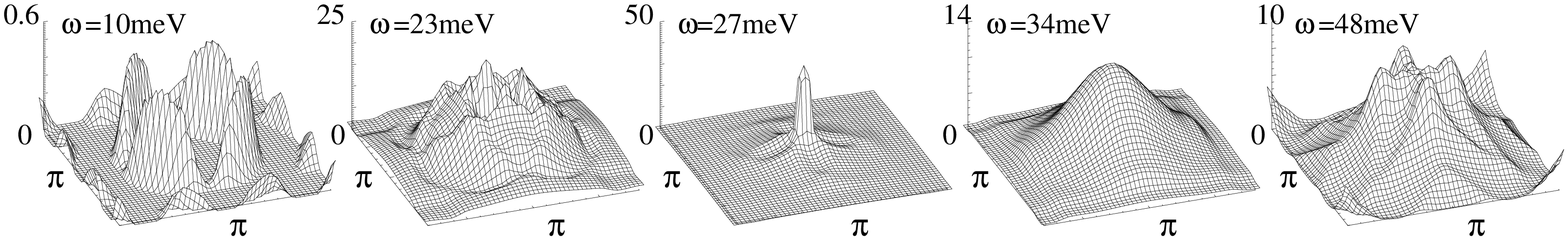}
}
\centerline{
\includegraphics[width=6.5in]{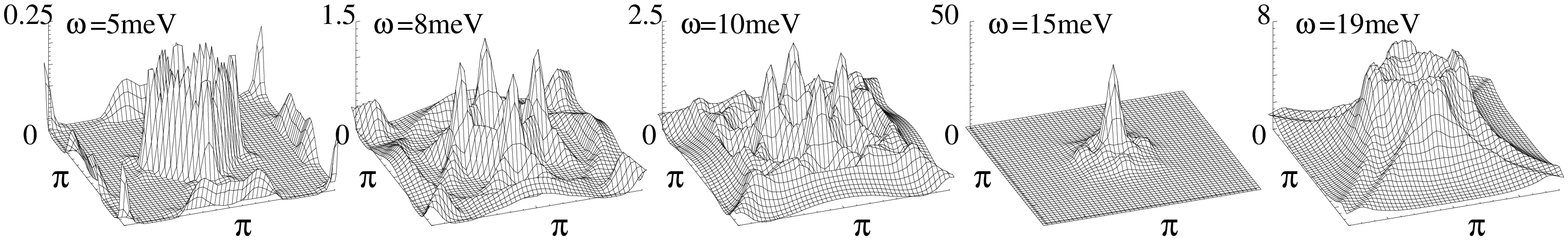}
}
\vskip 2mm
\caption{Frequency evolution of neutron cross section ($Im \chi$)
  for  YBaCuO$_{6.6}$ (upper panel) and  for optimal LaSrCuO (lower panel).} 
\end{figure*}

In Figure 1a we show the frequency evolution of  an YBaCuO$_{6.6}$
sample.  A node-to-node peak appears at low frequencies but the
magnitude is  small compared to that of all of the other features shown. 
Incommensurate peaks at $(\pi, \pi\pm\delta),(\pi\pm\delta,\pi)$ 
are first seen at around $\omega \approx \Delta$, albeit their magnitude 
is somewhat smaller than found experimentally.  
As frequency increases
there is a clear trend: the incommensurability decreases continuously;
this decrease is most apparent, in the immediate vicinity of the
resonance frequency. At $27 $ meV the resonant, commensurate peak is now well
established. As can be seen, there is some fine
structure near $(\pi , \pi) $ which is a remanent of the incommensurate
peaks. And there is a pronounced evolution in
the peak shape and height above resonance in the underdoped YBaCuO
sample. The $ (\pi , \pi) $ regime is a flat-topped, possibly weakly
incommensurate peak, just after resonance. It then broadens and remains
structureless (as shown by our $ 34 $ meV plot) between $ 34 - 48 $ meV.
Finally, above $ 48 $ meV, clear incommensurate structure appears.
 
Figure 1b shows the frequency evolution for the LaSrCuO family, here
shown at optimal doping. Node-to-node structures are seen at low frequencies
with very small amplitude. A four peaked structure is seen at
frequencies just around the gap frequency, $8 $ meV. These peaks are
sharper, but in roughly the same position as their counterparts in the
normal state\cite{Si}.  These peaks persist (with slightly growing
amplitude) until around $ 12 $ meV, at which point the
incommensurability seems to decrease while their overall magnitude
increases. In between $ 14 $ meV and $ 2 \Delta = 16 $ meV is an
interesting (evidently ``resonant") structure, but there have been, thus far, no
reports of this resonance. However, it should be stressed, that if
there is a commensurate feature in LaSrCuO, it should be seen only over a
very narrow ( $\approx 2 $ meV) frequency window. 
Finally, just beyond $ 2 \Delta $, as shown by the last panel at
$ \omega = 19 $ meV, the cross section becomes very similar in shape to
its normal state counterpart, although the magnitude is larger.

This evolution from incommensurate to commensurate peaks and then back to
incommensurate behavior with increasing $\omega $ can, thus, be seen to be
the case for both cuprates.  There are claims for these effects in recent
experiments on YBaCuO\cite{Japanese} and possibly in BiSSCO\cite{newMook}.
What appears to be different between
our observations and these particular experiments\cite{Japanese} is that we
do not find distinct two energy scales $E_c$, where the incommensurate peaks
merge and $E_r$, where the resonance occurs. Our calculations for the
reduced YBaCuO case, suggest that the incommensurate peaks probably never
merge, but rather that at resonance the $( \pi , \pi) $ feature fills in the
gap between the two incommensurate features. In this sense $ E_c$ and $E_r$
are the same frequency, although it should be stressed that here we have
incorporated no resolution limiting effects.

This work was supported by the NSF under awards No.~DMR-91-20000
(through STCS) and No.~DMR-9808595 (through MRSEC).

\end{document}